\documentclass[aps,pre,twocolumn,groupedaddress,showpacs,amsmath,eqsecnum]{revtex4}
   \usepackage{bm}
    \usepackage{graphicx}
\newcommand\be{\begin{equation}}
\newcommand\ee{\end{equation}}
\newcommand\ba{\begin{eqnarray}}
\newcommand\ea{\end{eqnarray}}
\newcommand{\nn}{\nonumber\\}
\newcommand{\av}[1]{\langle #1\rangle}  
\newcommand{\half}{\textstyle{\frac{1}{2}}}
\newcommand{\third}{\textstyle{\frac{1}{3}}}
\newcommand{\fourth}{\textstyle{\frac{1}{4}}}
\newcommand{\al}{{\alpha}}

\newcommand{\bv}{{\bf v}}

\begin{document}
\title{Exact steady state solution of the Boltzmann
equation:\\
 A driven 1-D inelastic Maxwell gas}

\author{A. Santos}
\email{andres@unex.es}
\affiliation{Departamento de F\'{\i}sica, Universidad de Extremadura,
E-06071 Badajoz, Spain}
\author{M. H. Ernst}
\email{ernst@phys.uu.nl} \affiliation{Instituut voor Theoretische
Fysica, Universiteit Utrecht, Postbus 80.195, 3508 TD Utrecht, The
Netherlands}
\date{\today}
\begin{abstract}
The exact nonequilibrium steady state solution of the nonlinear
Boltzmann equation for a driven inelastic Maxwell model was obtained
by Ben--Naim and Krapivsky [Phys. Rev. E \textbf{61}, R5 (2000)] in the
form of an infinite product for the Fourier transform of the
distribution function $f(c)$. In this paper we have inverted the
Fourier transform to express $f(c)$ in the form of an
infinite series of exponentially decaying terms. The
dominant high energy tail is exponential,  $f(c)\simeq
A_0\exp(-a|c|)$, where  $a\equiv 2/\sqrt{1-\alpha^2}$ and the
amplitude $A_0$ is given in terms of a converging sum. This is
explicitly shown  in the totally inelastic limit ($\alpha\to 0$)
 and in the quasi-elastic limit ($\alpha\to 1$).
In the latter case,  the distribution is dominated by a Maxwellian for a
very wide range of velocities, but  a crossover from a Maxwellian to an
exponential high energy tail  exists for velocities $|c-c_0|\sim 1/\sqrt{q}$
around a crossover velocity   $c_0\simeq \ln q^{-1}/\sqrt{q}$,   where
$q\equiv (1-\alpha)/2\ll 1$.
 In this crossover region the distribution function is extremely small,
$\ln f(c_0)\simeq q^{-1}\ln q$.

\end{abstract}
\maketitle
\section{Introduction\label{sec1}}

In kinetic theory there is a long standing interest in overpopulated
high energy tails of velocity distribution functions \cite{BKW}
because of chemical reactions and other activated processes that occur
only at energies far above thermal. This interest has been
considerably increased in the past 10 years because of research in
granular fluids with dissipative or inelastic interactions. The
velocity distributions in fluidized systems have been studied
theoretically \cite{WM96,W96,EP97,vNE98,SBCM98,BSSS99,BBRTW}, and measured
in Monte Carlo \cite{BBRTW,MC,MS00} and molecular dynamics simulations
\cite{vNETP99}, and in numerous laboratory experiments \cite{GM-exp}.

Very recently a revival in this field occurred when Baldassarri et
al.\ \cite{Rome1,Rome2} discovered an exact scaling solution --
with an algebraic high energy tail -- of the non-linear Boltzmann
equation for an inelastic one-dimensional freely cooling gas
(without energy input) with a collision frequency independent of
the energy of the colliding particles. This model, called
Inelastic Maxwell Model (IMM), was introduced by Ben--Naim and
Krapivsky \cite{BN-K-00}.  It is in fact an inelastic
modification of Ulam's stochastic model to illustrate the velocity
relaxation of elastic one-dimensional  point particles towards a
Maxwellian \cite{Ulam}. A three-dimensional version of it has been
constructed by Bobylev et al.\ \cite{Bobyl-00,Cercig-00}. For a
recent review  on inelastic Maxwell models, see Refs.\
\cite{EB-03,BNK03}.

Baldassarri et al.\ have demonstrated  the importance of this type
of solutions in \cite{Rome1} with the help of   Monte Carlo
simulations of the nonlinear Boltzmann equation for 1-dimensional
and 2-dimensional IMM's. It appeared that the solution, $F(v,t)$,
for large classes of initial distributions $F(v,0)$ (e.g. uniform
or Gaussian) and for all values of the inelasticity  could be
collapsed for large times on a scaling form $v_0^{-d}(t)
f(v/v_0(t))$,     where $v_0(t)=\langle v^2\rangle^{1/2}$ is the
r.m.s. velocity. In one dimension the scaling form was given by
$f(c) =(2/\pi)(1+c^2)^{-2}$, which has a heavily overpopulated
algebraic tail $\sim c^{-4}$ when compared to a
  Maxwellian.  In two dimensions the solutions also approached a scaling
form with an algebraic tail, $f(c) \sim c^{-d-a}$ with an exponent
$a(q)$ that depends on the degree of inelasticity $q= \half (1-\al)$,
 where $\al$ is the coefficient of restitution.
Soon after,  Ben--Naim and Krapivsky \cite{BN-K-02}, and
 Ernst and Brito \cite{B-E-02a}
obtained asymptotic solutions with algebraic tails for the velocity
distribution in $d$-dimensional freely cooling IMM's from
self-consistently determined solutions of the Boltzmann equation.
Using methods previously developed for   the  inelastic hard sphere
case, the asymptotic solutions were also extended to non-equilibrium
steady states (NESS) in $d$-dimensional systems driven by Gaussian
white noise and other thermostats \cite{vNE98,B-E-02c}. There the
tails exhibited over-populations of exponential type, $\sim
\exp(-a|c|)$, for all $d$-dimensional IMM's \cite{B-E-02c}. For
inelastic hard spheres, which is the prototypical model for granular
gases, the velocity distribution function  shows an overpopulated
exponential tail in free cooling \cite{vNE98,BBRTW,MC}, and a
stretched exponential tail $\sim \exp(-a|c|^{3/2})$, when driven by
white noise \cite{vNE98,BBRTW,MS00}.

For the case of $d$-dimensional free IMM's the approach of $F(v,t)$ to
a scaling form with an algebraic tail has also been  rigorously
proven, for initial distributions in the ${\cal L}_1$ function space,
satisfying the physical requirements of finite mass and energy, i.e.\
$\int d\bv \{1,v^2\} F(v,0) < \infty$ \cite{BCT02}.

What about exact and/or more explicit results for the distribution
function in the one-dimensional IMM, driven by Gaussian white noise?
The  exact solution of the nonlinear Boltzmann
equation for this case is given in the form of an infinite
product for the Fourier transform of the distribution function \cite{BN-K-00}.
 Nienhuis and van der Hart
\cite{Nienhuis} made an extensive numerical analysis of this solution,
and demonstrated exponential decay, in agreement with the predictions
of Ref.\ \cite{B-E-02c}. More numerical evidence for exponential high
energy tails in the one-dimensional driven IMM was given recently
by Marconi and Puglisi \cite{Rome3}, and  by  Antal et al.\
\cite{Droz}. In a recent paper \cite{BNK03}, Ben--Naim and Krapivsky
have also  used the Fourier
transform method to show  that the high energy tail is exponential
for any inelasticity, but with an amplitude that diverges in the
quasi-elastic limit. On the other hand, the problem of determining for
what range of velocities the exponential tail actually applies remains
open. This is one of the points addressed in this paper.

The plan of the paper is as follows. In the remainder of this Section
we present the nonlinear Boltzmann equation for the velocity
distribution function $F(v)$ or $f(c)$, driven by Gaussian white
noise, and we discuss qualitatively the physical properties of the
model in different limiting cases. In Section \ref{sec2} the exact
solution  $\phi(k)= \int dc\, e^{-ikc}f(c)$    of the Fourier
transformed Boltzmann equation in the NESS is presented in the form of
an infinite product, and its large- and small-$k$ properties are
analyzed. In Section \ref{sec3} we determine the inverse Fourier
transform, $f(c)$, in the form of an infinite series of exponentially
decaying terms. In the limit of totally inelastic collisions ($\al \to
0$) substantial simplifications occur. The rather singular
quasi-elastic limit ($\al \to 1$) is studied in Section \ref{sec4},
where also the crossover from   Maxwellian  to exponential decay is
analyzed. We end with some comments in Section \ref{sec5}, and some
technical details are moved to    Appendices \ref{appA} and
\ref{appB}.

 Before concluding this introduction we present the
Boltzmann equation for the one-dimensional inelastic Maxwell model
(IMM) \cite{BN-K-00}, driven by Gaussian white noise, and we discuss
some of its important properties. The time evolution of a spatially
homogeneous isotropic velocity distribution function $F(v,t)=F(|v|,t)$
is described by the nonlinear Boltzmann equation ,
\ba \label{a1}
&\frac{\partial F(v)}{\partial t} -D\frac{\partial^2 F(v)}{\partial
v^2} =\int dv_1\left[ \frac{1}{\al} F(v'')F(v_1'')-  F(v)F(v_1)
\right]&
\nonumber\\
&= -F(v) + \frac{1}{p}\int  du F(u)F\left(\frac{v-q u}{p}\right)
\equiv I(v|F).&
\ea
  All velocity integrations extend over the interval
$(-\infty,+\infty)$. The diffusion term represents the (heating)
effect of the Gaussian white noise with noise strength $D$. The
nonlinear collision term represents the inelastic collisions, where
$v''=v-\frac{1}{2}(1+\alpha^{-1})(v-v_1)$ and
$v_1''=v_1+\frac{1}{2}(1+\alpha^{-1})(v-v_1)$ denote restituting
velocities. Here $\al = 2p-1=1-2q$  with $0<\al<1$ is the coefficient
of restitution. The mass is normalized as $\int dv F(v)=1$, and the
mean square velocity or temperature as   $\av{v^2}(t)=\int dv v^2
F(v) \equiv v^2_0(t)$ . The rate equation,
\be \label{a2}
\partial_t \av{v^2} = 2D -2pq \av{v^2},
\ee
obtained from (\ref{a1}), describes the approach to the
non-equilibrium steady state (NESS)  with width $\av{v^2}
=D/pq$, where the heating rate $D$ caused by the random forces is
balanced by the loss rate, $pq \av{v^2}=
\frac{1}{4}(1-\al^2)\av{v^2}$, caused by the inelastic collisions.

 To understand the physical processes involved, we first
discuss in a qualitative way the relevant limiting cases. Without the
heating term $(D=0)$, Eq.\ (\ref{a1}) reduces to the freely
cooling IMM, whose exact solution has been discussed in Refs.
\cite{Rome1,Rome2}. If one takes in addition the elastic limit $(\al
\to 1$ or $q \to 0)$, the collision laws reduce in the {\it
one-dimensional} case to $v''=v_1,v''_1=v$, i.e.\ an exchange of
particle labels, the collision term vanishes identically, every
$F(v,t) = F(v)$ is a solution, there is no     randomization     or
relaxation of the velocity distribution through collisions, and the
model becomes trivial at the Boltzmann level of description, whereas
the distribution function in the presence of {\it infinitesimal}
dissipation $(\al \to 1)$ approaches a Maxwellian.

If we turn on the noise $(D \neq 0)$ at vanishing dissipation $(q=0)$,
the exact solution of (\ref{a1}) in Fourier representation is
$\widehat{F}(k,t) = \exp(-Dk^2t)\: \widehat{F}(k,0)$, and the granular
temperature, $v^2_0(t)=v^2_0(0)+2Dt$, increases linearly with time.
With stochastic heating {\it and} dissipation (even in infinitesimal
amounts) the system reaches a NESS, and it is the goal of this paper
to determine the NESS distribution function.

To expose the universality of this NESS it is convenient to measure
the velocities, $c =v/v_0(\infty)$, in units of its typical size
$v_0(\infty)$, i.e.\ the r.m.s. velocity or width of the velocity
distribution $v_0(\infty)$,
\be \label{a3}
F(v,\infty) = v_0^{-1}(\infty) f \left(v/v_0(\infty)\right),
\ee
which obeys the normalizations $\int dc \,\{1,c^2\} f(c)=\{1,1\}$.
Different normalizations have been used as well \cite{norm}.

The rescaled velocity distribution in the NESS is then the solution of
the scaling equation,
\be \label{b1}
I(c|f) = -\frac{D}{v_0^2(\infty)} f''(c) =-pq  f''(c),
\ee
where primes denote $c$-derivatives. The first equality may suggest
that $f(c)$  may depend on the noise strength $D$ and possibly on the
initial distribution via $v_0(\infty)$.  By eliminating $v_0(\infty)$
with the help of (\ref{a2}) in the NESS we have shown that the scaling
form of the distribution function $f(c)$ is a {\it universal}
function, that does not depend on  the strength $D$ of this thermostat,
nor on any property of the initial distribution. It only depends on
the type of thermostat used.

\section{Fourier transform of IMM Boltzmann equation\label{sec2}}
The nonlinear Boltzmann equation for characteristic function, $\phi(k)
=\int dc \, e^{-ikc} f(c)$, is obtained by Fourier transformation of
(\ref{b1}) with the result,
\be \label{b1bis}
(1+pqk^2) \phi(k) = \phi(pk)\phi(qk).
\ee
The simple structure of the equation for the Fourier transform
$\phi(k)$ follows because the nonlinear collision operator for
(in)elastic Maxwell models is a convolution in the velocity variables
\cite{BKW}.   Equation  (\ref{b1bis}) is a nonlinear finite difference
equation, that can be solved by iteration. A simple way to construct
the exact solution is to introduce $\psi(k)\equiv \ln \phi(k)$, which
satisfies
\begin{equation}\label{2}
\psi(k)=\psi(pk)+\psi(qk)-\ln\left(1+pqk^2\right).
\end{equation}
The normalization of mass and energy implies that $\phi(k)\approx
1-\frac{1}{2}k ^2$ and $\psi(k)\approx -\half k^2$ at small $k$. The
solution to (\ref{2})  can be found iteratively starting from
$\psi_0(k)=-\ln(1+pqk^2)$ and inserting $\psi_n(k)$ on the
right-hand-side of (\ref{2}) to get $\psi_{n+1}(k)$ on the
left-hand-side. By taking the limit
$\psi(k)=\lim_{n\to\infty}\psi_n(k)$, one finally obtains,
\ba
\psi(k)&=&-\sum_{m=0}^\infty\sum_{\ell=0}^m \nu_{m\ell}
\ln\left[1+p^{2\ell}q^{2(m-\ell)} pqk^2\right]
\nn \phi(k)&=&
\prod^\infty_{m=0}\prod^m_{\ell=0} [1+p^{2\ell}q^{2(m-\ell)}
pqk^2]^{-\nu_{m\ell}}, \label{3}
\ea
where $\nu_{m\ell}=\binom{m}{\ell} $. These solutions satisfy the
required boundary conditions at $k=0$. We further note that
$\overline{\psi}(k)= \psi(k) -\lambda|k|$ with $\lambda$ an arbitrary
complex number is also a solution of (\ref{2}), but in general  does
not satisfy the boundary conditions at small $k$. This property is a
reflection of the Galilean invariance of the original Boltzmann
equation.

Equations (\ref{3})  provide an exact representation in Fourier space
of the solution of the Boltzmann equation (\ref{a1}). The series
(\ref{3}) converges rapidly, even for large $k$. {} By expanding the
logarithm in powers of $k^2$ and summing a geometric series we obtain,
\begin{equation} \label{9}
\psi(k)=\sum_{n=1}^\infty
\frac{(-1)^n}{n}\frac{(k^2pq)^{n}}{1-p^{2n}-q^{2n}}.
\end{equation}
It  converges for  $k^2\leq 1/pq$, and $\psi (k)$ has  a branch point
singularity at $k^2=-1/pq$, as is apparent from (\ref{2}). Equation
(\ref{9}) allows one to get the cumulants $C_{2n}$
 defined by
\begin{equation} \label{9.1}
\psi(k)=\sum_{n=1}^\infty\frac{(-1)^n}{(2n)!}C_{2n}k^{2n},
\end{equation}
with the result
\begin{equation}\label{9.2.1}
C_{2n}=\frac{(2n)!}{n}\frac{(pq)^{n}}{1-p^{2n}-q^{2n}}.
\end{equation}
In particular,  $C_2=\langle c^2\rangle=1$. Since $1-p^{2n}-q^{2n}>0$,
it follows that {\it all} cumulants are positive, indicating already
an overpopulation of the high energy tails. So far a summary of the
results obtained in Ref. \cite{BN-K-00}.  We note that the Stirling
approximation shows that the cumulants  at fixed $\al$ or $q$ and $n >
e / (2\sqrt{pq})$ are rapidly diverging with increasing $n$, as
$C_{2n} \sim  2 \sqrt{\pi/n}(2n\sqrt{pq}/e)^{2n}$.

The exact solution $\phi(k)$ in (\ref{3}) has an infinite sequence of
poles of multiplicity $\nu_{m\ell}$ in the complex $k$-plane, all of
which contribute to the amplitude of the asymptotic high energy tail
of $f(c)$. This makes a numerical inversion of $\phi(k)$ to obtain
$f(c)$ a bit tricky. To determine $f(c)$ several authors
\cite{Nienhuis,Droz}   have performed numerical inversions of
$\phi(k)$, starting from the infinite product (\ref{3}) or from the
more convenient  series expansion (\ref{9}). However the latter one is
only convergent for $ pqk^2< 1$. To facilitate such numerical
procedures, we have derived an expansion in powers of $k^{-2}$,
convergent in the complementary region, $pqk^2> 1$, of the complex
$k$-plane.  This rather technical part is deferred to Appendix
\ref{appA}. The results can be found in (\ref{4}), (\ref{8}), and
(\ref{6}).

\section{High energy tail \label{sec3}}
\begin{figure}[tbp]
\includegraphics[width=.90 \columnwidth]{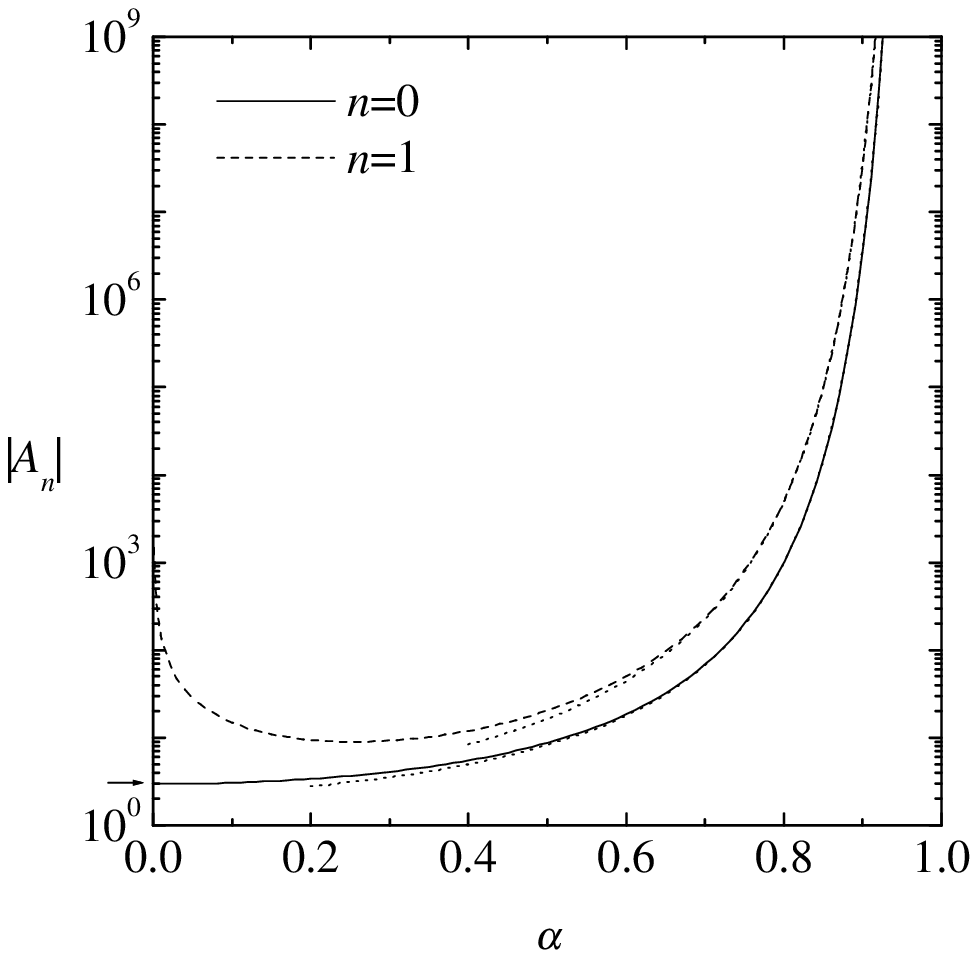}
\caption{Logarithmic plot of the amplitudes $A_0\equiv A_{000}$ (solid
line) and $-A_1\equiv -A_{110}$ (dashed line) as functions of the
coefficient of restitution. The arrow indicates the value $A_0\simeq
2.958389$ at $\alpha=0$. The dotted lines represent  the asymptotic
form (\protect\ref{c25}) for small $q$. \label{fig1}}
\end{figure}

\begin{figure}[tbp]
\includegraphics[width=.90 \columnwidth]{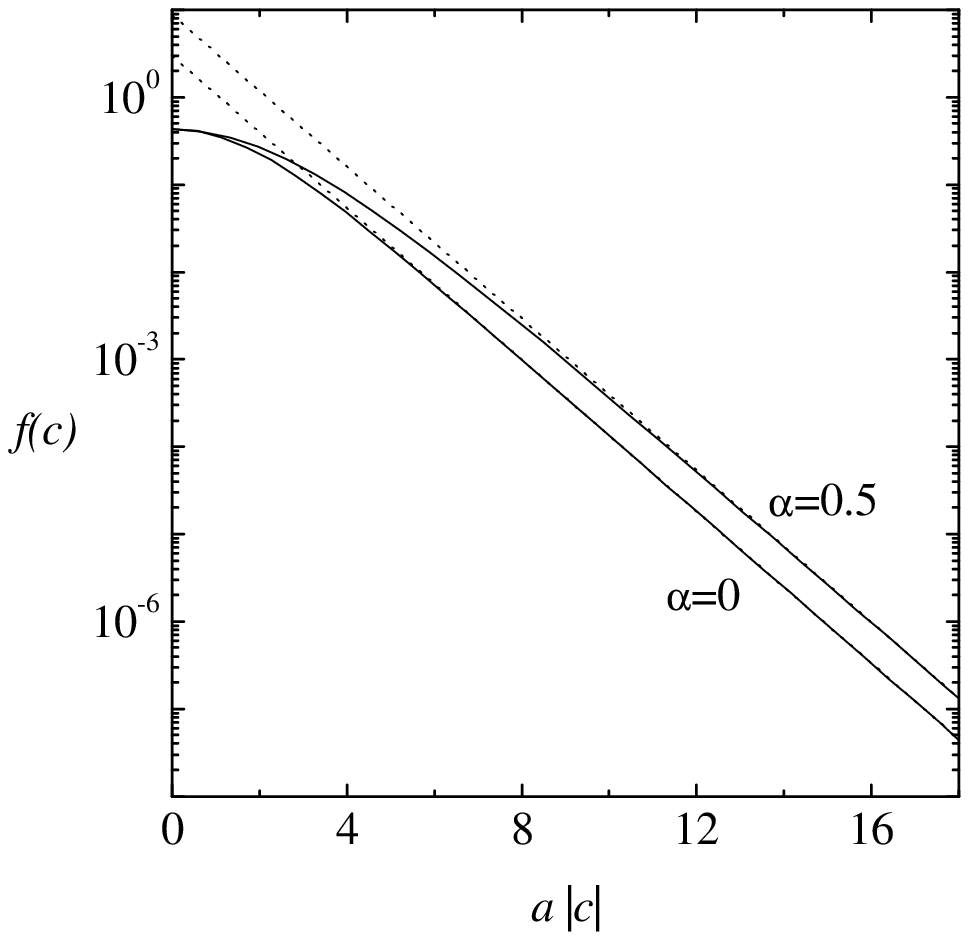}
\caption{Logarithmic plot of $f(c)$ versus $a|c|$ for  $\alpha=0$ and
$\alpha=0.5$. The dotted lines are the asymptotic forms $f(c)\approx
A_0 e^{-a|c|}$ at $\alpha=0$ and $\alpha=0.5$, with $A_0$ obtained from
(\ref{c7}). \label{fig2}}
\end{figure}

On account of (\ref{3}), the characteristic function $\phi(k)$ can be
written as
\begin{equation}\label{c1}
\phi(k)=\prod_{m=0}^\infty\prod_{\ell=0}^m
\left(1+k^2/k_{m\ell}^2\right)^{-\nu_{m\ell}},
\end{equation}
where $k_{m\ell}\equiv ap^{-\ell}q^{-(m-\ell)}$ with $a\equiv
1/\sqrt{pq}$. Thus $\phi(k)$ has poles at $k=\pm i k_{m\ell}$ with
multiplicity $\nu_{m\ell}$. The velocity distribution,
\begin{equation}\label{c2}
f(c)=\frac{1}{2\pi}\int_{-\infty}^\infty dk\, e^{i kc}\phi(k),
\end{equation}
can then be obtained by contour integration. As $f(c)$ is an even
function, we only need to evaluate the integral in (\ref{c2}) for
$c>0$. The replacement $c\to|c|$ gives then the result for all $c$. By
closing the contour through an infinite upper half-circle and applying
the residue theorem we obtain
\begin{equation}\label{c3}
f(c)=\sum_{m=0}^\infty\sum_{\ell=0}^m
e^{-k_{m\ell}|c|}\sum_{n=0}^{\nu_{m\ell}-1}|c|^n A_{m\ell n},
\end{equation}
where
\ba \label{c4}
A_{m\ell
n}&=&\frac{i^{n+1}k_{m\ell}^{2\nu_{m\ell}}}{n!(\nu_{m\ell}-1-n)!}\lim_{k\to
ik_{m\ell}} \left(\frac{\partial}{\partial k}\right)^{\nu_{m\ell}-1-n}
\nonumber\\
&&\times
(k+ik_{m\ell})^{-\nu_{m\ell}}\widetilde{\phi}_{m\ell}(k),
\ea
with
\begin{equation}\label{c5}
\widetilde{\phi}_{m\ell}(k)\equiv\prod_{m'=0}^\infty\prod_{\ell'=0}^{m'}
\left(1+k^2/k_{m'\ell'}^2\right)^{-\nu_{m'\ell'}\left(1-\delta_{m
m'}\delta_{\ell \ell'}\right)}.
\end{equation}
 Note that the factor labeled $(m',\ell') = (m,\ell)$ is absent. The
dominant terms in (\ref{c3}) for large $|c|$ correspond to the
smallest values of $k_{m\ell}$. The two smallest ones are
$k_{00}=a$ and $k_{11}=a/p$. Consequently, the leading and
sub-leading terms are
\begin{equation}\label{c6}
f(c)\approx A_0e^{-a|c|}+A_1e^{-a|c|/p}+\cdots,
\end{equation}
where
\be \label{Am}
A_n \equiv A_{nn0} =(a/2p^n) \widetilde{\phi}_{nn}(i a /p^n)\vspace{3mm} .
\ee
We calculate the first two explicitly, i.e.
\ba  \label{c7}
A_0&=& \frac{a}{2} \exp \left[\sum_{m=1}^\infty
\frac{p^{2m}+q^{2m}}{m \ (1- p^{2m}-q^{2m})}\right]\\
A_1&=& \frac{- ap^3}{2(1-p^2)(p-q)}\nonumber\\
&&\times \exp\left[\sum_{m=1}^\infty\frac{p^{-2m}(p^{2m}+q^{2m})^2}{m\
(1- p^{2m}-q^{2m})}\right]
\label{c7bis}.
\ea
In the last equalities we have followed steps similar to those used to
obtain (\ref{9}) from (\ref{3}). The results  (\ref{c3})--(\ref{c7bis})
exhibit the full analytic structure of the dominant and subdominant
high energy tails of the velocity distribution in the NESS, as already
demonstrated numerically for the one-dimensional case in Refs.\
\cite{Nienhuis,Rome3,Droz},   and derived in \cite{B-E-02c} for
$d$-dimensional IMM's on the basis of self-consistent solutions.
Moreover we have obtained here explicit expressions for the amplitudes
$A_0$ and $A_1$ in the form of sums that are rapidly converging when
$q$ is not too small.  The coefficients     $A_0\equiv A_{000} $ and
$A_1\equiv A_{110}$ are shown in Fig.\ \ref{fig1} as  functions of
$\al$,    where $A_{110} \propto 1/(p-q) = 1/\al$ diverges according
to  (\ref{c7bis}).
 The next term to those explicitly given in (\ref{c6}) corresponds
either to $k_{22}=a/p^2$ if $p^2>q$ (i.e., if $\alpha>\sqrt{5}-2\simeq
0.236$) or to $k_{10}=a/q$ if $p^2<q$. Note that the amplitude
$A_{100}$ of $\exp(-k_{10}|c|)$ can be obtained from $A_1$ in
 (\ref{c7bis})  by interchanging $p\leftrightarrow q$.
    Figure \ref{fig2} compares the asymptotic form $f(c)\approx A_0
e^{-a|c|}$ with the function $f(c)$ obtained by numerically inverting
$\phi(k)$ for $\alpha=0$ and $\alpha=0.5$. We observe that the
asymptotic behavior is  reached for $a|c|\gtrsim 4$ if $\alpha=0$
and for $a|c|\gtrsim 8$ if $\alpha=0.5$. As $a=1/\sqrt{pq}$ this
corresponds to velocities far above the r.m.s. velocity.

There are two interesting limiting cases: the {\it quasi-elastic}
limit $(\al \to 1,  q \to 0)$,  and the {\it totally inelastic}  limit
$(\al \to 0, p \to \half^+, q \to \half^- )$. We start with the
latter.  In the totally inelastic limit $(\al \to 0)$ the subdominant
terms $A_{110} e^{-a|c|/p}$ and $A_{100} e^{-a|c|/q}$ become equally
important, i.e.\ the single poles in (\ref{c1}) at $k_{11}=a/p$ and
$k_{10}=a/q$ coalesce, and (\ref{c6}) no longer describes the
subdominant large-$c$ behavior correctly. Moreover $A_{110}\simeq -
A_{100} \propto 1/\al$, as can be seen in Fig.\ \ref{fig1} for
$A_{110}$. In fact, the poles $k_{m\ell} \to k_m \equiv 2^m a$
coalesce for all $\ell$, some of the coefficients $A_{m\ell n}$
diverge, e.g. $A_{mm0} \propto (1/\al)^{2^m-1}$,
and the expansion     makes no sense anymore. So, we analyze the case
$\al =0$ separately. In this case the characteristic function is
according to (\ref{3}),
\begin{equation} \label{c11}
\phi(k)=\prod_{m=0}^\infty \left(1+k^2/k_{m}^2\right)^{-\nu_{m}},
\end{equation}
where $\nu_{m}\equiv 2^m$ and $k_{m}\equiv 2^{m}a$ with
$a=1/\sqrt{pq}=2$. Then the distribution function is
\begin{equation} \label{c13}
f(c)=\sum_{m=0}^\infty
e^{-k_m|c|}\sum_{n=0}^{\nu_{m}-1}|c|^n A_{m n},
\end{equation}
where the residues or amplitudes are given by,
\ba \label{c14}
A_{mn}&=&\frac{i^{n+1}k_{m}^{2\nu_{m}}}{n!(\nu_{m}-1-n)!} \lim_{k\to
ik_{m}} \left(\frac{\partial}{\partial k}\right)^{\nu_{m}-1-n}
\nonumber \\
&&\times (k+ik_{m})^{-\nu_{m}}\widetilde{\phi}_{m}(k),
\ea
and $\widetilde{\phi}_n(k)$ is defined as,
\begin{equation} \label{c15}
\widetilde{\phi}_{n}(k)\equiv\prod_{m=0}^\infty
\left(1+k^2/k_{m}^2\right)^{-\nu_{m}\left(1-\delta_{n m}\right)}.
\end{equation}
For large $|c|$ the distribution function becomes,
\begin{equation} \label{c16}
f(c)\approx A_{00}e^{-2|c|}+(A_{10}+A_{11}|c|)e^{-4|c|}+\cdots.
\end{equation}
To calculate the amplitudes of the dominant terms we derive from
(\ref{c15}),
\ba \label{c17}
 \ln \widetilde{\phi}_{0}(k)&=& -\sum_{m=1}^\infty 2^m \ln (1+2^{-2m}
k^2/a^2)
\nn &=& \sum_{n=1}^\infty
\frac{(-1)^n}{n}\frac{(k/a)^{2n}}{2^{2n-1} -1}
\nn \ln \widetilde{\phi}_{1}(k)&=& - \ln(1+k^2/a^2)
\nn & & + 2 \sum_{n=1}^\infty
\frac{(-1)^n}{n}\frac{(k/2a)^{2n}}{2^{2n-1} -1}.
\ea
The definitions  (\ref{c14})--(\ref{c15}) with $k_m = 2^m a \:(a=2)$
yield then,
\ba \label{c18}
A_{00} &=& \half a \widetilde{\phi}_{0}(ia)=
e^{S_0} \simeq 2.958389
\nn A_{11} &=&  a^2 \widetilde{\phi}_{1}(2ia)=
-\textstyle{\frac{4}{3}}A^2_{00} \simeq -11.669422
\nn A_{10}&=& \half
a \widetilde{\phi}_{1}(2ia)-ia^2 \widetilde{\phi}^\prime_{1}(2ia)
\nn &=& \textstyle{\frac{4}{3}}\left(S_1 -\textstyle{\frac{11}{12}}\right)
A^2_{00}\simeq
3.138267,
\ea
where we have used the rapidly converging sums,
\ba \label{c19}
S_0 &=& \sum_{n=1}^\infty \frac{1}{n} (2^{2n-1} - 1)^{-1} \simeq
1.084645
\nn S_1&=&  \sum_{n=1}^\infty  (2^{2n-1} - 1)^{-1} \simeq
1.185597.
\ea
In fact, the results (\ref{c16}) could have been derived
directly from  (\ref{c3})--(\ref{c7bis})  after lengthy calculations, by
expanding $A_{110}$ and $A_{100}$ in powers of $\al$, with the result,
\be \label{c20bis}
A_{1s0} = (-1)^{s+1} A_{11}/(8 \al) +\half A_{10} +{\cal O}(\al)
\ee
with $s=0,1$. Insertion of these results in (\ref{c3}) yields
(\ref{c16}). The limit $\al \to 1$ is discussed in the next
Section.

\section{Quasi-elastic limit\label{sec4}}

As already mentioned in the introduction, the velocity distribution
approaches a NESS, even in the presence of an \textit{infinitesimal}
dissipation ($\al \to 1, q \to 0$), balanced by a ditto amount of
stochastic heating.  This limit is referred to as the quasi-elastic
limit.  For the rescaled functions $f(c)$  and $\phi (k)$ it  simply
refers to the limit $q \to 0$.

Once we have \textit{first} taken the large $|c|$ limit at
\textit{fixed} $\alpha<1$ --- as has been done in the previous Section
--- we can \textit{next} take the quasi-elastic limit $\alpha\to 1$.
When the limits are taken in that order, the asymptotic behavior is
still of the form $e^{-a|c|}$, where the decay constants   are  $k_{mm}=
a/p^m \to a$, and the amplitudes may diverge. On the other hand, if
the limits are taken in the reverse order, \textit{first} $\alpha\to
1$ at \textit{fixed} $|c|$ and \textit{next} $|c|\to \infty$, the
behavior is in general totally different.

 First consider the second case, and observe that $\psi(k)$
in (\ref{9}) has at small $q$ the form $ \psi(k) = -\half k^2 +
\sum_{n=2}^\infty a_{2n}(q) k^{2n}$ with rapidly decreasing
coefficients $a_{2n} \simeq(-1)^n q^{n-1} (1-\half q)/(2n^2)$ for $n
\geq 2$. Consequently $\phi(k) = e^{\psi(k)}$ can be expanded as,
\be \label{d1}
\phi(k) = e^{-\half k^2} \left[ 1+\sum_{n=2}^\infty \mu_{2n}(q) k^{2n}
\right],
\ee
where the relation between  $a_{2n}$ and $\mu_{2n}$ is the same as
between cumulants and moments after setting $a_2=\mu_2=0$. The
coefficients $\mu_{2n}$ are to dominant order in $q^2$ given by
\ba \label{d2}
\mu_4 &=& a_4 =\textstyle{\frac{1}{8}}q (1-\half q- \frac{3}{4} q^2
)+ {\cal O}(q^4)
\nn \mu_6 &=& a_6 =- \textstyle{\frac{1}{18}} q^2 (1- \half q)+ {\cal
O}(q^4)
\nn \mu_8 &= & \half a^2_4+a_8 =\textstyle{\frac{1}{128}} q^2
(1+3q)+{\cal O}(q^4)
\nn \mu_{10} &\simeq & a_4 a_6 =-\textstyle{\frac{1}{144}}q^3 +{\cal O}(q^4)
\nn \mu_{12} &\simeq & \textstyle{\frac{1}{6}} a^3_4 =
\frac{1}{3072} q^3 +{\cal O}(q^4),
\ea
and in general $\mu_{4n-2} \sim \mu_{4n} \sim {\cal O}(q^n)$ for
$n\geq 2$.  The series above can be Fourier inverted term-wise,
using the following relation,
\ba \label{d3}
&\int^\infty_{-\infty} \frac{dk}{2 \pi} e^{ikc} e^{-\half k^2} k^{2n}
= (-1)^n\left(\frac{d}{dc}\right)^{2n} \exp(-\half c^2)/\sqrt{2\pi}&
\nn &=  (-1)^n H\!e_{2n}(c) f_0(c)
= 2^n n! L_n^{(-1/2)} (\half c^2)f_0(c),&
\ea
where $f_0(c)= \exp(-\half c^2)/\sqrt{2\pi}$ is the Maxwellian. In the
last two equalities Rodrigues' formula for the Hermite polynomials has
been used, as well as their relation to the generalized Laguerre or
Sonine polynomials (see Ref. \cite{A+S}, Eqs.\ (22.11.88), (22.5.18) and
(22.5.40)). The resulting Sonine polynomial expansion of the velocity
distribution in the NESS reads then,
\be \label{d4}
f(c)=f_0(c) \left[1+\sum_{n=2}^\infty (-1)^n \mu_{2n}(q)
H\!e_{2n}(c)\right].
\ee
Similar expansions of the NESS-distribution function in low order
Hermite or Sonine polynomials have also been derived for inelastic
hard spheres in $d$-dimensions \cite{vNE98}, and for a
3-dimensional IMM in \cite{Cercig-00}.

Next we consider the case, where \textit{first} $|c| \to \infty$ at
finite $\al<1$,  and \textit{next} $\al \to 1$, or $q \to 0$. The
large-$c$ behavior at fixed $\al$ has already been discussed in
 (\ref{c3})--(\ref{c7bis}),  and we observe that the terms
 in Eq.~(\ref{c3}) at
large $c$, associated with  all poles of the form
$k_{n\ell}=a/p^{\ell}q^{n-\ell}\:(\ell<n)$ decay rapidly as $q \to 0$, and only poles with $k_{nn}=a/p^n$ need to be considered:
\be
f(c)=\sum_{n=0}^\infty A_n e^{-k_{nn}|c|}.
\label{new1}
\ee
We will analyze the behavior of the associated amplitudes $A_{n}$  by
combining (\ref{Am}) with (\ref{c5}), i.e.
\ba \label{c20}
&\ln A_n =  \ln (a/2p^n) & \nn &-\sum_{m=0}^\infty\sum_{\ell=0}^m \binom{m}{\ell}
(1-\delta_{mn}\delta_{\ell n})  \ln\:[ 1-p^{-2(n-\ell)}q^{2(m-\ell)}]&
\nn &
\equiv B_n^{(1)} + B_n^{(2)} + B_n^{(3)} +\ln( a/2p^n) \:,&
\ea
 where
\ba \label{c21} B_n^{(1)}& =&  - \sum_{m=0}^{n-1} \sum_{\ell=0}^{m}
\binom{m}{\ell} \ln \:[ 1-p^{-2(n-\ell)}q^{2(m-\ell)}]
\nn
 B_n^{(2)} &=& -\sum_{\ell=0}^{n-1} \binom{n}{\ell}
 \ln\:[ 1-(q/p)^{2(n-\ell)}]
\nn B_n^{(3)} &=& - \sum_{m=n+1}^\infty \sum_{\ell=0}^{m} \binom{m}{\ell}
\ln \:[ 1-p^{-2(n-\ell)}q^{2(m-\ell)}] .\nn
&&
\ea
Now we take the limit $q \to 0$ at {\it finite} $n$ and retain terms
to order $q$.  The
dominant small-$q$ contribution to $B_n^{(1)}$ comes from $\ell =m$,
i.e.
\ba \label{c21a}
B_n^{(1)}& =& -\sum^n_{m=1} \ln \:(1-p^{-2m}) +o(q)
\nn  &=& - \ln \:[(-2q)^n n!]-\half n(n+2)q+o(q),
\ea
where we used the relation $1-1/p^{2m} \simeq -2mq [1+(m+\half)q]$,
and $o(q^k)$ denotes terms which are negligible with respect to $q^k$.
Note that the complex number $B_n^{(1)}$ is only determined modulo $\{
2\pi i\}$, but $\exp\left(B_n^{(1)}\right)$ is single-valued. Furthermore, we
observe that $B_n^{(2)} = {\cal O}(n q^2)$. The analysis of
$B_n^{(3)}$ in (\ref{c21}) is more involved and given in Appendix \ref{appB}.
The result is,
\be \label{c24}
B_n^{(3)} = \frac{\pi^2}{12q} +\half \ln q  -K_0 +\half \left(n +
\frac{13}{12} - \frac{\pi^2}{72}\right)q+o(q),
\ee
where
\be \label{c24bis}
K_0 = \frac{3}{4} +\frac{\pi^2}{24} -\half \ln 2 -R \simeq
 0.733598.
\ee
Combining the small-$q$ results (\ref{c21a}) and (\ref{c24}) for
$B_n^{(1)}$
 and $B_n^{(3)}$ with (\ref{c20}) yields for $A_n$,
\ba \label{c25}
&A_n= \frac{a}{2p^n} \exp\:[B_n^{(1)}+ B_n^{(3)}  +o(q)]& \nn &=
\frac{(-1)^n}{2 n! (2q)^n}  \exp\left[\frac{\pi^2}{12q} -K_0
 - \half n(n-1)q+K_1 q+o(q)\right]& ,\nn
\ea where
\be
K_1 = \frac{25}{24} - \frac{\pi^2}{144} \simeq 0.9731278.
\ee

 To describe the crossover between the two different
limiting behaviors, i.e.\ (\ref{d4}) with first $q \to 0$, next $c \to
\infty$, and (\ref{new1}), (\ref{c25}) with first $c \to \infty $, next
$q \to 0$ we need to couple these limits, which will be discussed
next.

By an extension of the steps followed in Appendix \ref{appB}, it can
be verified that the terms denoted by $o(q)$ in Eq.~(\ref{c25})
have the form $n^{k_1}q^{k_2}$ with $k_1\leq k_2+1$ and $k_2\geq 2$.
Therefore, those terms can be neglected against the terms of order $q$
if $n\ll q^{-1}$.

\begin{figure}[tbp]
\includegraphics[width=.90 \columnwidth]{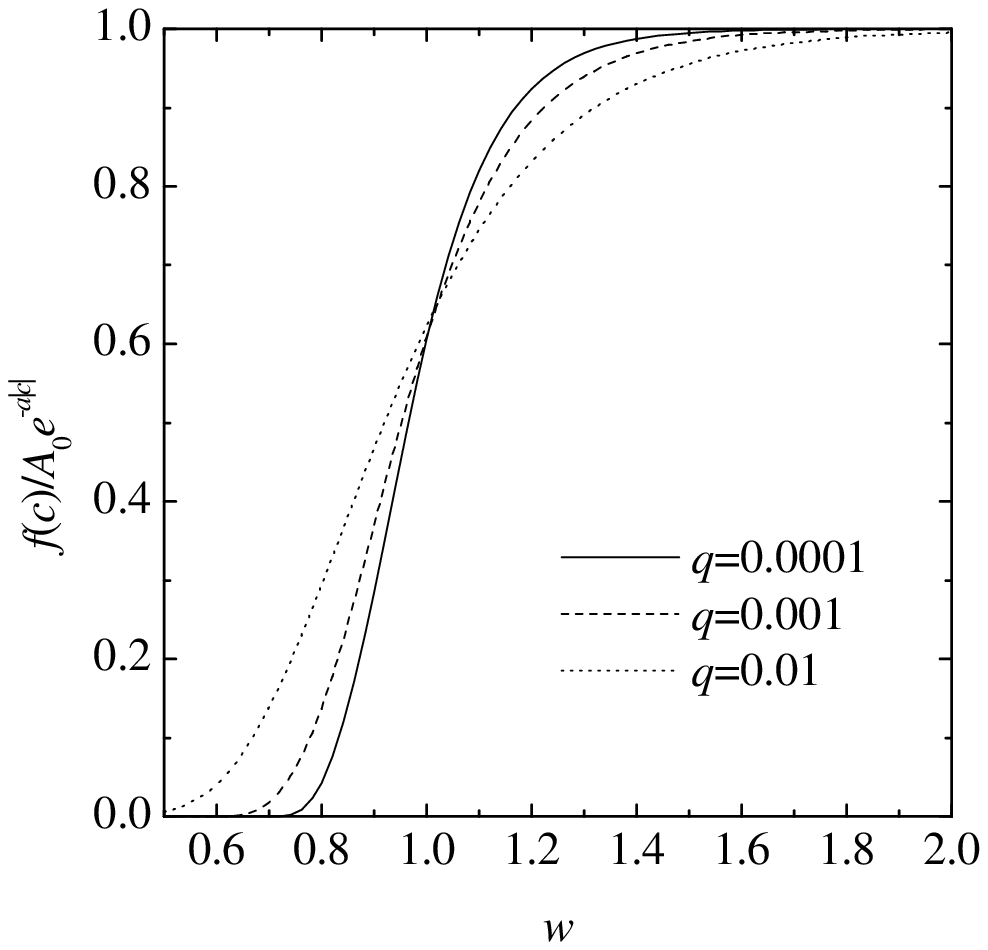}
\caption{Plot of the ratio between the velocity distribution function
$f(c)$ and its  high energy tail $A_0e^{-a|c|}$ as a function of the
scaled velocity $w\equiv |c|\sqrt{q}/\ln q^{-1}$ in the quasi-elastic
limit for $q=0.01$, 0.001, and 0.0001. \label{fig4}}
\end{figure}

The ratio  $R(c)$  between the distribution
function $f(c)$ in (\ref{new1}) and its asymptotic high energy
form $A_0 e^{-a|c|}$, defines a \textit{crossover} function
\be
R(c)\equiv f(c)/A_0 e^{-a|c|}=\sum_{n=0}^\infty b_n r_n, \label{new2}
\ee
where $r_n$ and $b_n=A_n/A_0$  follow from (\ref{new1}) and
(\ref{c25}) as,
\ba
r_n &=& \exp\left[-a|c|(p^{-n}-1)\right] \nonumber  \\
b_n&=& \frac{(-1)^n}{n!(2q)^n}\exp\left[-\half n(n-1)q+o(n^2q)\right].
\label{new3}
\ea
 Here we have written $o(q)\to o(n^2q)$ to emphasize the fact that
Eq.~(\ref{new3}) remains valid if $n\ll q^{-1}$. So there is a
  crossover   behavior in $R(c)$ from a large-$c$ behavior of   ${\cal
O}(\exp[-c^2/2])\simeq 0$   in the small-$q$ Sonine polynomial expansion
(\ref{c4}), to the the small-$q$ behavior of $R(c)$ of ${\cal O}(1)$
in (\ref{new2}). The transition region is characterized by a
crossover velocity $c_0$ such that $R(c_0)\approx \half$. The
interesting questions are, how does $c_0$ scale with $q$ in the
quasi-elastic limit, and what is the width of the crossover
region? To address these questions, note that the series (\ref{new2})
converges for all velocities and the signs of the terms are
alternating. Therefore, when breaking off the infinite sum at $n=N$,
the maximum error is $|b_{N+1}|r_{N+1}$:
\ba
R(c)&=&\sum_{n=0}^N b_n r_n+\Delta^{(N)}(c)\nn &\equiv&
 R^{(N)}(c)+\Delta^{(N)}(c),\quad |\Delta^{(N)}(c)|\leq
|b_{N+1}|r_{N+1}.\nn && \label{new4}
\ea
  This suggests that the pure exponential high energy tail $A_0 e^{-a|c|}$
 qualitatively describe the large-$c$ behavior of $f(c)$ if
\be
|b_1| r_1=\frac{e^{-a|c|q/p}}{2q}\simeq \frac{e^{-\sqrt{q}|c|}}{2q}\leq \half.
\label{new5}
\ee
Of course the bound $\half$ may be replaced by any number of the order
of 1 in this estimate. Equation (\ref{new5}) implies that   $w\equiv |c|
\sqrt{q}/\ln q^{-1}\geq 1$.   Therefore, we can estimate the
crossover velocity to be $c_0=(\ln q^{-1})/\sqrt{q}$ or, equivalently,
$w_0=1$. To confirm this and get a closed form for the crossover
function $R(c)$, consider a value of $w$ in the range $0.5<w<1$ and
take $N=\beta q^{w-1}$, where $\beta\gtrsim 1$. In that case, $N\gg 1$
but $N^2 q\ll 1$, so that $r_n\simeq q^{nw}$ and $b_n\simeq
(-1)^n/n!(2q)^n$ for $n\lesssim N$, and $|b_{N+1}|r_{N+1}\simeq
(2\beta/e)^{-N}/2\beta\sqrt{2\pi N}$. Therefore, with this choice of
$N$,
\ba
R^{(N)}(c)&\simeq &\sum_{n=0}^N \frac{(-1)^n}{n!}
\left(\frac{q^{w-1}}{2}\right)^n,\nn |\Delta^{(N)}(c)|&\leq&
\frac{1}{2\beta\sqrt{2\pi N}}\left(\frac{2\beta}{e}\right)^{-N}.
\label{new6}
\ea
If $\beta>e/2\simeq 1.36$ then $\Delta^{(N)}(c)\ll 1$ and $R(c)$ can
be approximated by $R^{(N)}(c)$. By the same arguments,  the upper
limit in the summation of (\ref{new6}) can be replaced by infinity.
The choice $N=\beta q^{w-1}$ is justified by the fact that for $w<1$
the term $|b_n|r_n$ reaches a high maximum value
$|b_{n_0}|r_{n_0}\simeq \exp(n_0+\half)/\sqrt{2\pi n_0}$ at $n_0\simeq
\half (q^{w-1}-1)$ and then decays rapidly. If $w>1$, however,
$|b_n|r_n$ decreases monotonically and thus $\Delta^{(N)}(c)\ll 1$
for any choice of $N$. In conclusion, the crossover function for
$w>0.5$ in the quasi-elastic limit becomes
\be
R(c)\simeq \exp\left(-q^{w-1}/2\right), \quad w\equiv |c|\sqrt{q}/\ln
q^{-1}. \label{new7}
\ee
At $w=1$ we have $R(c=c_0)\simeq 1/\sqrt{e}\simeq 0.6$, thus
confirming the estimate of the crossover velocity $c_0$ made below
Eq.~(\ref{new5}). Figure \ref{fig4} represents the crossover function
$R(c)$ versus the scaled velocity $w$ for $q=0.01$, 0.001, and 0.0001.
To measure the width of the crossover region, let  $w_1$ and $w_2$
denote  the values of $w$ at which $R=0.1$  and $R=0.9$,
respectively. From Eq.~(\ref{new7}) we obtain $w_1\simeq 1-1.5/\ln
q^{-1}$, $w_2\simeq 1+1.6/\ln q^{-1}$, so the width scales as $w_2-w_1
\sim 1/\ln q^{-1}$. Going back to unscaled velocities, the crossover
takes place between $c_1=c_0-1.5/\sqrt{q}$ and
$c_2=c_0+1.6/\sqrt{q}$ with a width $c_2-c_1\sim 1/\sqrt{q}$.
For $q=0.01$, 0.001, 0.0001 one has $c_0\simeq 46$, 218, 921 and
$c_2-c_1\simeq 31$, 98, 310, respectively. For these high values of
the velocity the distribution function is extremely small. For
instance, at $c=c_0$,   $f(c_0)\simeq \half \exp\left[(q^{-1}+\half)\ln
q+\pi^2/12q -K_0-\half\right]$.   This yields $f(c_0)\sim
10^{-166},10^{-2645}, 10^{-36431}$ for $q=0.01$, 0.001, 0.0001,
respectively. These values are beyond the accuracy of any numerical or
simulation method, so the high energy tail in the quasi-elastic limit
would look like a Maxwellian for the domain of velocities numerically
accessible. On the other hand, our asymptotic analysis of the exact
solution shows that the true tail is actually exponential.

\section{Conclusion\label{sec5}}
The exact nonequilibrium steady state solution of the nonlinear
 Boltzmann equation for a driven one-dimensional inelastic Maxwell
 gas was obtained in Ref.~\cite{BN-K-00} in the form of an infinite
product for the Fourier transform $\phi(k)$ of the distribution
function $f(c)$. The main goal of this paper has been to show that
this relatively simple exact solution in the one-dimensional case
also possesses the generic properties of overpopulation of high
energy tails and exhibits a rich mathematical structure, especially in
the different limiting cases.

We have inverted the Fourier transform to express $f(c)$ in the form
of an infinite series of exponentially decaying terms, as given by
Eq.~(\ref{c3}) with the velocity $c$ measured in units of the r.m.s
velocity (i.e.\ $\langle c^2\rangle^{1/2}=1$). For all values of the
coefficient of restitution $0\leq\alpha<1$ the high energy tail is
exponential, namely $f(c)\simeq A_0\exp(-a|c|)$, where $a\equiv
1/\sqrt{pq}=2/\sqrt{1-\alpha^2}$ and the amplitude $A_0$ is given by
Eq.~(\ref{c7}) and plotted in Fig.~\ref{fig1}.

Special attention has been paid to two complementary limiting cases:
the totally inelastic limit ($\alpha\to 0$) and the quasi-elastic
limit ($\alpha\to 1$). In the former case some poles coalesce and
the dominant high energy term is still exponential, but  the
subdominant term becomes an exponential times a  linear function of
the velocity, where the numerical value of the associated
amplitudes is given by (\ref{c18}).

The quasi-elastic limit is much more delicate and requires some care.
If we first take $\alpha\to 1$ at fixed $|c|$  and next
$|c|\to\infty$ (order A), the high energy tail has a Maxwellian form. On the
other hand, if the limits are taken in the reverse order, i.e.\ first
$|c|\to\infty$ at fixed $\alpha<1$ and then $\alpha\to 1$ (order B), the
asymptotic high energy tail is exponential.  The crossover between
both limiting behaviors is described by the coupled limit $c \to
\infty$ and $q \to 0$ with the scaling variable   $w = |c| \sqrt{q}/\ln
q^{-1} =\text{fixed}$   with   $q \equiv \half (1-\alpha)\ll 1$,   and occurs at $w
\simeq 1$.   If $w<1$ (more specifically, $1-w\gtrsim 1/\ln
q^{-1})$, the distribution function is essentially a Maxwellian, while
the true exponential high energy tail is reached if $w>1$ (more
specifically, $w-1\gtrsim 1/\ln q^{-1}$).

\begin{table*}[t]
\caption{\label{table1}
Asymptotic behavior of the distribution function $f(c)$ for one-dimensional systems
in the quasi-elastic limit. In general, the result depends on the order of limits. Order A corresponds to take first $\alpha\to 1^-$ and then $|c|\to \infty$, whereas order B refers to the reverse order, i.e. first $|c|\to \infty$ and then $\alpha\to 1^-$. The first/second footnote in the second column gives the reference where the result for order A/B was obtained.}
\begin{ruledtabular}
\begin{tabular}{llcc}
State&System &Order A&
Order B\\
\hline
Free cooling&Hard spheres\footnotemark[1]\footnotemark[2]&$\frac{1}{2}\left[\delta(c-1)+\delta(c+1)\right]$&$e^{-a|c|}$
\\
 &Maxwell model\footnotemark[3]\footnotemark[3]&$c^{-4}$&$c^{-4}$\\
White noise&Hard spheres\footnotemark[1]\footnotemark[4]&$e^{-a|c|^3}$&$e^{-a|c|^{3/2}}$\\
&Maxwell model\footnotemark[5]\footnotemark[6]&$e^{-ac^2}$&$e^{-a|c|}$\\
Gravity thermostat&Hard spheres\footnotemark[5]\footnotemark[7]&$\frac{1}{2}\left[\delta(c-1)+\delta(c+1)\right]$& $e^{-ac^2}$ \\
&Maxwell model\footnotemark[5]\footnotemark[8]&$\frac{1}{2}\left[\delta(c-1)+\delta(c+1)\right]$& $e^{-a|c|}$\\
\end{tabular}
\end{ruledtabular}
\footnotetext[1]{Ref.\ \protect\cite{BBRTW}}
\footnotetext[2]{Ref.\ \protect\cite{vNE98}}
\footnotetext[3]{Refs.\ \protect\cite{Rome1,BN-K-00,B-E-02a}}
\footnotetext[4]{Refs.\ \protect\cite{vNE98,BBRTW}}
\footnotetext[5]{This work}
\footnotetext[6]{Refs.\ \protect\cite{BBRTW,B-E-02c}}
\footnotetext[7]{Ref.\ \protect\cite{MS00}}
\footnotetext[8]{Ref.\ \protect\cite{B-E-02c}}
\end{table*}
It is of interest to emphasize that the results for the scaling form
in the quasi-elastic limit not only depend sensitively on the order in
which both limits are taken. They also depend strongly on the
collisional interaction, i.e. on the energy dependence of the
collisional frequency, as well as on the mode of energy supply to the
system.
To illustrate this we have collected in Table \ref{table1} what is known for the
different inelastic models in one dimension: (i) hard spheres and (ii)
Maxwell models, and for different modes of energy supply: (i) no
energy input or free cooling, (ii) energy input or driving through
Gaussian white noise, represented by the forcing term $-D \partial^2
F(v,t)/\partial v^2$ in the Boltzmann equation, and (iii) energy  input
through a {\it negative} friction force $\propto g v/|v|$, acting in
the direction of the particle's velocity, but independent of its
speed. This driving, referred to as gravity thermostat, can be
represented as the forcing term $ g (\partial /\partial v)[ (v/|v|) F(v,t)]$ in the
Boltzmann equation.
The results corresponding to order A with the gravity thermostat have been obtained by the same method as followed in Ref.\ \cite{BBRTW}.
 It is worthwhile noting that in the quasi-elastic
limit a bimodal distribution, $\frac{1}{2}\left[\delta (c+1)+\delta (c-1)\right]$, is
observed in inelastic hard sphere systems both for free cooling and
for driving through the gravity thermostat, whereas in inelastic
Maxwell models this bimodal distribution is only observed for the
gravity thermostat.

It is important to note that in the normalization where velocities
are measured in units of the r.m.s. velocity, the high energy tail in
the driven inelastic Maxwell model is only observable for very large
velocities, as illustrated in Figure \ref{fig2} for strong $(\al \to
0)$ and intermediate $(\al =\half)$ inelasticity. In the quasi-elastic
limit, where $(\al \to 1)$, the tail is even pushed further out
towards infinity, as analyzed at the end of Section \ref{sec4}. This
also explains how to reconcile the paradoxical results of
exponential large-$c$ behavior with the very accurate
representation (\ref{d4}) of the distribution function  in the
thermal range, in the form of a Maxwellian, multiplied by a
polynomial expansion in Hermite or Sonine polynomials with
coefficients related to the cumulants. The validity of these
polynomial expansions, over a large range of inelasticities with $(0
\leq \al <1)$ had been observed before in \cite{vNE98} for inelastic
hard spheres and in \cite{Cercig-00} for inelastic Maxwell models. On
the other hand the high energy tail is $\propto e^{-a|c|}$, and not
$\propto c^N e^{-c^2/2}$, where $N$ is some large number, and yields
diverging moments   $M_{2n}=\langle c^{2n}\rangle$ and cumulants
$C_{2n}$    in the limit $n\to\infty$, as shown in Section \ref{sec2}.

The exact solutions of the nonlinear Boltzmann equation for the freely
evolving \cite{Rome1} and the driven \cite{BN-K-00} inelastic Maxwell
model (extended in this paper), as well as the rigorous proof of
\cite{BCT02} for the long time approach of the distribution function
to a scaling form, validate the self consistent method, developed in
\cite{vNE98} for analytical studies of possible over- or
underpopulations of the high energy tail of velocity distributions,
not only for inelastic Maxwell models, but -- more importantly -- also
for inelastic hard sphere, where exact solutions are not known. This
possibility of assessing the validity of general kinetic theory
methods by means of exact solutions of the nonlinear Boltzmann
equation is one of the main reasons why the study of inelastic Maxwell
models is of interest.

\acknowledgments
The authors are indebted to B. Nienhuis for
 stimulating  discussions about the subject of this paper.
  A.S. acknowledges partial support from the Ministerio de
Ciencia y Tecnolog\'{\i}a
 (Spain) through grant No.\ BFM2001-0718.

\appendix
\section{Large-${k}$ Expansion\label{appA}}

\begin{figure}[tbp]
\includegraphics[width=.90 \columnwidth]{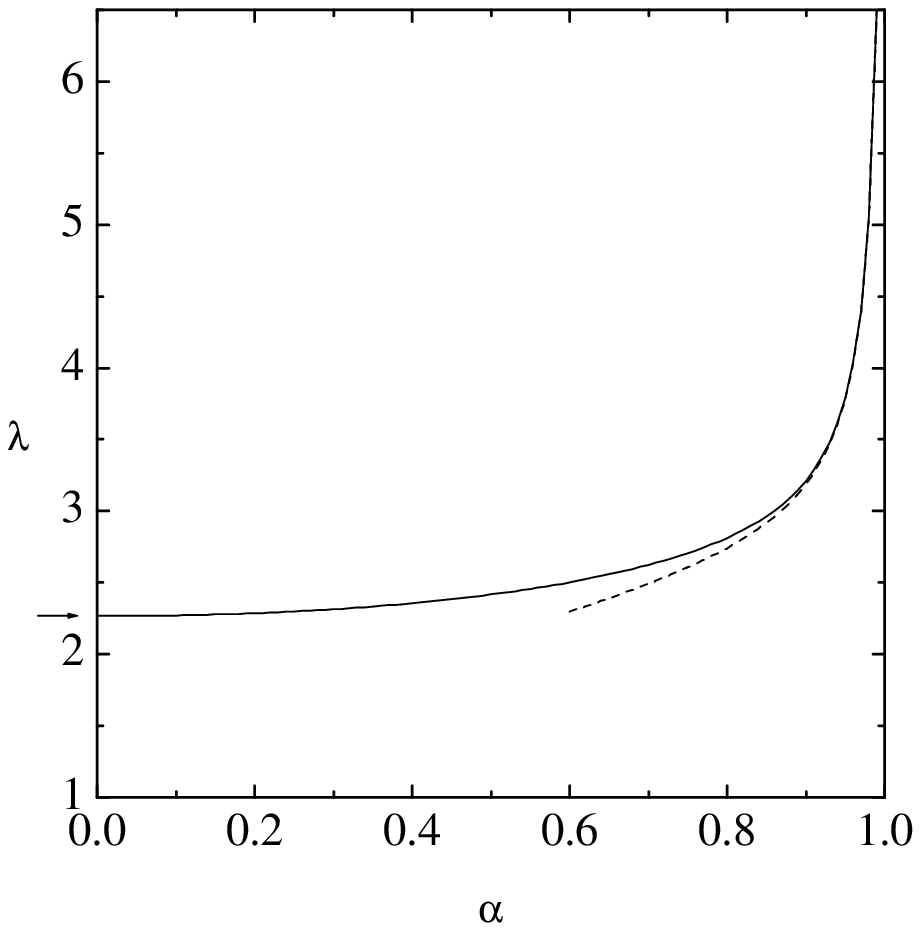}
\caption{Coefficient $\lambda$ as a function of the coefficient of
restitution. The arrow indicates the value $\lambda=\pi/2\ln 2$ at
$\alpha=0$. The dotted line represents  the asymptotic form Eq.~(\protect\ref{6}) for small
$q$. \label{fig5}}
\end{figure}

The asymptotic behavior of $\psi$ for large $k$  can be obtained by
inserting the ansatz, $\psi=- \lambda|k|+\ln (Ak^2)+ \sum^\infty_{n=1}
a_n k^{-2n}$, with unknown coefficients $\{A,a_n\}$ into (\ref{2}),
and equating the coefficients of equal powers of $\ln k$ and $k^{n}$
with the result,
\be  \label{9.3}
\psi(k) =-\lambda|k|+\ln (k^2/pq) -\sum_{n=1}^\infty
\frac{(-1)^n}{n}\frac{(k^2pq)^{-n}}{p^{-2n}+q^{-2n }-1},
\ee
where $\lambda$ is as yet undetermined. The series converges for
$k^2\geq q/p$. However for the $\psi(k)$ above to qualify as a
solution of (\ref{2}) the radius of convergence is further restricted
to $(qk)^2\geq q/p$ or $pq k^2\geq 1$. The constant $\lambda$ must be
chosen such that $\psi(k)$ satisfies the boundary condition $\psi
\simeq -\half k^2$ at small $k$. This can be done by matching
(\ref{9.3}) with (\ref{9}). The latter satisfies already the small-$k$
boundary condition. Matching at $pq k^2 =1$ yields then,
\ba
\label{4} &\frac{\lambda}{\sqrt{pq}} =\:-\:2 \ln(pq) +
\sum^\infty_{n=1} \frac{(-1)^{n+1}}{n} &
\nn & \times \left(\frac{1}{1-p^{2n}-
q^{2n}}+ \frac{1}{p^{-2n}+q^{-2n }-1}\right).&
\ea
Both terms can be combined into a single $n$-sum with $n= \pm 1,\pm 2,
\cdots$. The above result is not only convenient for numerical
evaluation, as shown in Fig.~\ref{fig5}, but also for analytic
evaluation in two limiting cases.
We first consider the {\it totally inelastic} limit ($\al \to 0$ or $
p=q=\half $). There the expansion (\ref{9.3}) can be cast into a
simpler form,
\ba \label{7}
\psi(k) &=&-\lambda|k|+ \ln (4k^2) - \half {\sum_{n=1}^\infty}
\frac{(-1)^n}{n}\frac{k^{-2n}}{1-2^{-(2n+1) }}
\nn &=&-\lambda|k| +\ln (4k^2)+\half {\sum^\infty_{m=0}}\frac{1}{
2^{m}} \ln(1+\frac{1}{2^{2m} k^{2}})
\nn &=&-\lambda|k| +\half \sum^\infty_{m=0} \frac{1}{2^{m}}
\ln\:(1+2^{2m} k^2).
\ea
Matching this expression in $k^2= 1/pq=4$ with the exact solution in
(\ref{3}), $ \psi(2) = -\sum^\infty_{m=0}2^m \ln\:( 1+2^{-2m})$,
yields the nice result,
 \ba \label{8}
\lambda &=& \half \sum^\infty_{m=-\infty} 2^m \ln(1+2^{-2m})
\nn &=& \frac{1}{4 \ln 2} \int^\infty_0 dx x^{-3/2} \ln (1+x) =
\frac{\pi}{2 \ln 2}.
\ea
One can verify using the Euler--MacLaurin summation formula (see
 Eq. (23.1.30) of \cite{A+S}) that all correction terms to
the integral are vanishing, and the integral is listed in Eq.
(4.293.3) of \cite{Grad-Ryz}.

In the {\it quasi-elastic} limit $(\al \to 1$ or $q \to 0$) the sum
originating from the second term inside $( \cdots)$ in (\ref{4}) is of
${\cal O}(q^2)$, and will be neglected. To evaluate the first term $T$
for small $q$ we expand it as follows,
\ba \label{9bis}
T &=&\sum_{k=1}^\infty \frac{(-1)^{k+1}}{k}
\frac{1}{1-p^{2k}}\left[1+ \frac{q^{2k}}{1-p^{2k} } + {\cal
O}(q^{2(2k-1)}) \right]
\nn &=& \sum_{k=1}^\infty\frac{(-1)^{k+1}}{k}\left[\frac{1}{2kq}
+\frac{2k-1}{4k} +
\frac{q^{2k}}{(2kq)^2} +{\cal O}(q)  \right] \nn
&=& -\frac{1}{2q} \mbox{Li}_2(-1) -\half \mbox{Li}_1(-1) +\fourth
\mbox{Li}_2(-1) +\fourth + {\cal O}(q),\nn
&&
\ea
where  the polylogarithmic functions are defined as,
\be \label{polylog}
\mbox{Li}_k(x) = \sum^\infty_{n=1} {x^n}/{n^k}
\ee
 with $\mbox{Li}_2(-1)= -\frac{1}{12} \pi^2$ and
$\mbox{Li}_1(-1)=-\ln 2$~\cite{Wolfram}. The final result for
$\lambda$ at small $q$ is then,
\ba \label{6}
\lambda &=& \frac{1}{\sqrt{q}}
\left[ \frac{\pi^2}{24} - 2q \ln q +\half q(\ln 2 +\half -
\frac{\pi^2}{12})\right. \nn
 & &\left.  +q^2 \ln q + {\cal O}(q^2)
\right].
\ea

\section{Asymptotics in quasi-elastic limit}
\label{appB}

To calculate $B_n^{(3)}$ in (\ref{c21}) for small $q$ we expand the
logarithm, and perform the $(m,\ell)$-summation. The result is,
\ba \label{B1} 
&B_n^{(3)} = \sum_{k=1}^\infty \frac{p^{2k}}{k(1-p^{2k})}
\frac{(1+q^{2k}/p^{2k})^{\: n+1}}{1-q^{2k}/(1-p^{2k})}&
\nn &= \sum_{k=1}^\infty \frac{p^{2k}}{k(1-p^{2k})} \left[1+
\frac{q^{2k}}{1-p^{2k}} + \frac{q^{4k}}{(1-p^{2k})^2} + \frac{(n+1)
q^{2k}}{p^{2k}} \right] +o(q)& \nn &\equiv S(x) +\delta S(n,x) +o(q), &
\ea
where all contributions $\propto q^0$ and $\propto q$ have been
included. The dominant term is,
\be \label{B2}
S(x)  = \sum_{k=1}^\infty \frac{e^{-kx}}{k(1-e^{-kx})} \qquad (x=-2
\ln p).
\ee
In the remaining contributions to (\ref{B1}) only the term $k=1$ needs
to be taken into account, and yields
\be \label{B3}
\delta S(x,n) = \fourth +\half q (n+\frac{3}{4}) \simeq \fourth
+\fourth x (n+\frac{3}{4}).
\ee

To study the small-$x$ behavior of (\ref{B2}) we construct an
asymptotic series for $S(x)$, by expanding $1/(1-e^{-kx})$ in powers
of $x$. This can be done most conveniently by using the small-$x$
expansion of $x \coth x$, or equivalently the generating function for
the Bernoulli numbers, $B_{2k} =(-1)^{k+1} |B_{2k}|$ (see Eqs
(23.1.1-2) of \cite{A+S}), which we write as,
\be \label{B4} 
 \frac{1}{1-e^{-x}}= \frac{1}{x} +\frac{1}{2} + \sum_{m=1}^\infty
\frac{B_{2m}}{(2m)!} x^{2m-1}.
\ee
Substitution of (\ref{B4}) with $x \to kx$ into (\ref{B2}) yields,
\ba
\label{B5} S(x) &=& \frac{1}{x}\mbox{Li}_2(e^{-x}) -\frac{1}{2} \ln
(1-e^{-x})
 \nn &+& \sum^\infty_{m=1} \frac{ x^{2m-1}}{(2m)!}B_{2m}
\mbox{Li}_{2-2m}(e^{-x}).
\ea
We have used the definition (\ref{polylog}) of the polylogarithmic
functions, which are all singular in $x=0$. To determine the behavior
of the dilogarithm, $\mbox{Li}_2(e^{-x})$ we use the functional
relation (see Eq.\ (5) of \cite{Wolfram}),
\ba \label{B6} 
&\mbox{Li}_2(e^{-x}) =\frac{\pi^2}{6} - \ln(e^{-x}) \ln(1-e^{-x}) -
\mbox{Li}_2(1-e^{-x})&
\nn &= \frac{\pi^2}{6} + x (\ln x - \half x) -(x - \fourth x^2)
+{\cal O}(x^3).&
\ea
Here the small-$x$ expansion of the sum in (\ref{B5}) can be obtained
from the relation,
\ba \label{B7}
&\mbox{Li}_{-n}(e^{-x}) = \sum_{k=1}^\infty k^n e^{-kx}&
 \nn &=
\left( - \frac{d}{dx}\right)^n \mbox{Li}_0(e^{-x}) = \left( -
\frac{d}{dx}\right)^n ( e^x-1)^{-1}&
\nn &= \left( -\frac{d}{dx} \right)^n \left\{ \frac{1}{x} - \frac{1}{2}
+ \sum^\infty_{m=1}\frac{B_{2m} }{(2m)!} x^{2m-1}\right\}&
\nn &= \frac{n!}{x^{n+1}} \left[ 1- \half x \delta_{n0} +o(x)
\right]. &\ea
 The small-$x$ expansion of $1/(e^x-1)$ has been obtained
from (\ref{B4}) with $x \to -x$.

By combining the relations (\ref{B6}) and (\ref{B7}) with the
small-$x$ expansion of $\ln (1-e^{-x})$, we obtain from (\ref{B5}),
\ba \label{B8}
S(x) &=& \frac{\pi^2}{6x} + \half \ln (1-e^{-x}) -
\frac{1}{x}\mbox{Li}_{2}(1-e^{-x})
\nn &&+ \sum^\infty_{m=1}\frac{B_{2m} }{2m (2m-1)} (1-\half x \delta_{m1})
\nn &=& \frac{\pi^2}{6x} + \half \ln x -1 +R_0 - \frac{x}{24} +O(x^2),
\ea
  where   $R_0 \equiv \sum^\infty_{m=1} B_{2m}/[2m(2m-1)]$.   As  $|B_{2k}|
\sim 2 (2k)!/(2 \pi)^{2k}$ for $k \to \infty $ (see Eqs.\ (23.2.16)
and (23.2.18) of \cite{A+S}), the series is a divergent asymptotic
series with {\it alternating } signs. One obtains the greatest
accuracy, denoted by   $R_0^{(m_0)}$,   if one breaks off the series just
before the smallest term in the series, which is defined to be the
$(m_0+1)$-th term. Then the maximum error is $|B_{2 m_0+2} /(2
m_0+1)(2m_0+2))|$ (see Chap. 0.33 of \cite{Grad-Ryz}). In the present
case one can simply verify that $m_0=3$, and the best possible
estimate for the remainder in the limit where $q \to 0$ is given by,
\be \label{B9} 
R_0 =R_0^{(3)} \pm \frac{1}{56}|B_8| \simeq 0.0813492 \pm 0.0005952.
\ee
  The inaccuracy in $S(x)$, caused by the inaccuracy in the asymptotic
series   $R_0$,   can be substantially  reduced -- if so desired --  by
restricting the $m$-sum in (\ref{B5}) to $m_0=3$ terms, and
calculating the difference,
\ba \label{B10}
 \Delta(x) &=& \sum_{k=1}^\infty \frac{e^{-kx}}{k} \left[ \frac{1}{1-e^{-kx}} -
\frac{1}{k x} -\frac{1}{2} \right.  \nn  & &-   \left.
 \sum^3_{m=1} \frac{B_{2m} }{(2m)!} (kx)^{2m-1} \right],
\ea
in the small-$x$ limit as an integral. The result for instance, to
seven decimal points is $\Delta = - 0.0002877$. Hence
\be \label{B11}
R_0=R_0^{(3)} +\Delta \simeq   0.0810615.
\ee
  Combination of the results (\ref{B1}), (\ref{B3}) and (\ref{B8}) gives
the dominant small-$x$ behavior of $B_n^{(3)}$ in the form,
\be
B_n^{(3)} = \frac{\pi^2}{6x} + \half \ln x -\frac{3}{4} +R_0 +  \fourth
x (n + \frac{7}{12}).
\ee
  Final elimination of $x= 2q (1+\half q +\third q^2 + \cdots )$ in
favor of $q$ gives (\ref{c24}) in the main text.


\begin{thebibliography}{}
\bibitem{norm}  The normalization chosen here is the same as in
\cite{Rome1}.
In Refs.\   \cite{B-E-02a,B-E-02c,Rome3}   the normalization is $\av{c^2}= 1/2$ and
$D=\half pq$ when specialized to a one-dimensional system, and in
Refs.\ \cite{Nienhuis,Droz} one has $\av{c^2}=1/pq$ and $D=1$.
\end{thebibliography}
\end{document}